\documentclass{basi}
\usepackage[british]{babel}
\usepackage[varg]{txfonts}
\begin{document}

\title[Transient High Mass X-ray Binaries]{Transient High Mass X-ray Binaries}

\author[B.~Paul and S.~Naik]{Biswajit Paul$^1$\thanks{email:
       \texttt{bpaul@rri.res.in}} and Sachindra Naik$^2$\thanks{email:
       \texttt{snaik@prl.res.in}}\\
       $^1$Raman Research Institute, Sadashivnagar, C.~V.~Raman Avenue,
           Bangalore 560080, India\\
       $^2$Physical Research Laboratory, Navrangapura,
           Ahmedabad 380009, Gujarat, India}

\pubyear{2011}
\volume{39}
\pagerange{\pageref{firstpage}--\pageref{lastpage}}

\date{Received 2011 October 7; accepted 2011 October 19}

\maketitle

\label{firstpage}

\begin{abstract}
High Mass X-ray Binaries (HMXBs) are interesting objects that provide a wide
range of observational probes to the nature of the two stellar components,
accretion process, stellar wind and orbital parameters of the systems. A large
fraction of the transient HMXBs are found to be Be/X-ray binaries in which the
companion Be star with its circumstellar disk governs the outburst. These
outbursts are understood to be due to the sudden enhanced mass accretion to the
neutron star and is likely to be associated with changes in the circumstellar
disk of the companion. In the recent years, another class of transient HMXBs
have been found which have supergiant companions and show shorter bursts.
X-ray, infrared and optical observations of these objects provide vital
information regarding these systems. Here we review some key observational
properties of the transient HMXBs and also discuss some important recent
developments from studies of this class of sources. The X-ray properties of these
objects are discussed in some detail whereas the optical and infrared
properties are briefly discussed.
\end{abstract}

\begin{keywords}
 stars: binaries: general -- stars: pulsars: general -- stars: neutron -- X-rays: binaries -- 
 X-rays: bursts -- transients 
\end{keywords}

\section{Introduction}

X-ray binaries are the brightest X-ray sources in the sky. In these systems, a
compact object which is either a white dwarf, a neutron star or a black hole,
and a normal star which is in the process of evolution, revolve around the
common centre of mass. Depending on the mass of the binary companion, these
systems are classified into two types such as (a) Low Mass X-ray Binaries
(LMXBs; mass of the companion $\leq 3$ M$_\odot$) and (b) High Mass X-ray
Binaries (HMXBs; mass of the companion $\geq 10$ M$_\odot$). The HMXBs contain
massive and early-type (O or B-type) companion stars where as in case of LMXBs,
the spectral type of the binary companion is A-type or later. It is seen that
the HMXBs are generally concentrated towards the Galactic plane in contrast to
the LMXBs which are found near  the Galactic centre, Galactic bulge, in the
Galactic plane, and also in the globular clusters in the halo. Mass-accretion
takes place from the binary companion to the compact object through Roche-lobe
overflow (in case of LMXBs) or/and capture of stellar wind of the companion (in
case of HMXBs). As the binary companion evolves and fills the Roche-lobe, mass
transfer takes place from the companion star to the compact object through the
inner Lagrange point. In case of most HMXBs, however, the dominant process of
mass accretion from the companion to the compact object is by the capture of
the stellar wind of the companion. The accreted mass then spirals around the
compact object forming a disk like structure, known as an accretion disk, before
falling on to the compact object. The gravitational pull of the neutron star or
black hole in the binary system accelerates the matter to extremely high
velocities. The accelerated matter approaches the compact star, and if it has a
stellar surface, most of the kinetic energy of the accreted material is
converted to thermal energy which is radiated away primarily in the X-ray
energy band. In some cases, especially if the compact star does not have a strong
magnetic field, there can also be a outflow in the form of a collimated jet
from the compact star or a disk wind.


The HMXB systems are strong X-ray emitters by the process of accretion of
matter from the OB companion. Based on the type of the companion star, these
systems are further classified into Be/X-ray binaries and supergiant X-ray
binaries. The Be/X-ray binaries represent the largest subclass of high mass
X-ray binaries. About 2/3 of the identified systems fall into this category. In
these systems, the optical counterpart of the neutron star is either a dwarf,
subgiant or a giant OB star (luminosity class of III, IV or V) which shows
spectral lines in emission. The mass donor in the Be binary systems is
generally a B star that is still on the main sequence and lying well inside the
Roche surface. In these Be binary systems, the compact object (almost always a
neutron star) is typically in a wide orbit with moderate eccentricity with
orbital period in the range of 16--400 days. Evolutionary calculations show
that Be star and white dwarf or Be star and black hole should also be common
types of systems. However, no clear evidence of existence of such systems has
been shown as yet (Zhang, Li \& Wang 2004 and references therein). Mass
transfer from the Be companion to the neutron star takes place through the
equatorial circumstellar disc thought to be formed from the matter expelled
from the rapidly rotating Be star. The neutron star in these systems spends
most of the time far away from the circumstellar disc surrounding the Be
companion. The X-ray spectra of these Be/X-ray binaries are usually hard. The
hard X-ray spectrum along with the transient nature are important
characteristics of the Be/X-ray binaries.

In case of supergiant X-ray binaries, the optical counterpart is a star of
luminosity class of I or II. In these systems, the compact object orbits around
the supergiant early-type star which is deep inside the highly supersonic wind.
The optical companion emits a substantial stellar wind with a mass loss rate of
10$^{-6}$--10$^{-8}$ M$_\odot$ yr$^{-1}$ with a terminal velocity up to 2000
km~s$^{-1}$. The supergiant X-ray binaries are further subdivided into two
groups according to the dominant mode of mass transfer: (a) capture of the
stellar wind of the optical companion and (b) Roche lobe overflow. In some of
these systems, both types of mass transfer may be taking place (Blondin,
Stevens \& Kallman 1991). Although capture from a high-velocity stellar wind is
inefficient, the large mass-loss rate in the wind can result in an appreciable
mass accretion rate onto the neutron star that is sufficient to emit radiation
in X-ray band. Vela~X-1 is the well known wind-fed supergiant X-ray binary
pulsar. In case of Roche lobe overflow, the mass donor fills its Roche lobe and
results in transfer of material from the companion to the neutron star through
the first Lagrange point and forms an accretion disk around the neutron star.
This is a very efficient form of accretion and results in a mass transfer rate
much larger than by capture of the wind alone. The X-ray photons emitted from
the compact object through either of the accretion processes, must propagate
through the stellar wind to the observer, which causes absorption, scattering
and reprocessing of the X-ray spectrum. Some of the supergiant X-ray binaries
in which the Roche lobe overflow is the dominant mode of mass accretion, are
usually persistent X-ray sources. SMC~X-1, LMC~X-4 and Cen~X-3 are amongst the 
best candidates for the disk-fed (via Roche lobe overflow) supergiant X-ray binary
pulsars.


Most of the high mass X-ray binaries are transient sources that are usually
quiescent and occasionally become X-ray bright for a duration of a few days to
several tens or hundreds of days. During the peak of the outburst, some of
these sources are among the brightest X-ray sources in the sky. During the
X-ray outbursts, the X-ray brightness/luminosity of some of the transient
sources increases by up to four to five orders of magnitudes. The huge range of
luminosity change makes these objects interesting to study in X-ray bands.
While the X-ray bursts, outbursts, flares are generally transient and
unpredictable, the Be/X-ray transients often show regular and periodic
outbursts on time scales in the range of tens of days to several hundreds of
days, same as their orbital periods.

\begin{figure}
\centering
\includegraphics[height=5.3in, width=1.5in, angle=-90]{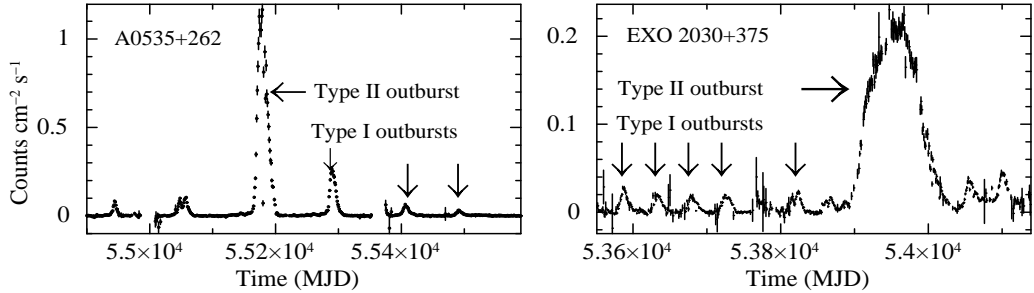}
\caption{The outburst profiles of Be/X-ray binaries A0535$+$262 (left panel) and
EXO~2030$+$375 (right panel) are shown in 15--50 keV energy band, as seen by
\textit{Swift}/BAT. Type~I and Type~II X-ray outbursts seen in these systems are
marked in both panels.}\label{X-ray_outburst}
\end{figure}

The Be/X-ray binaries display two types of X-ray outbursts such as Type~I and
Type~II outbursts. Type~I outbursts are short and periodic outbursts of
moderate luminosity ($L_{\rm X} \leq 10^{35}{-}10^{37}$ erg~s$^{-1}$) which
occur close to the time of periastron passage of the neutron star in the wide
eccentric orbit. This suggests an enhanced accretion caused by the proximity of
the neutron star to the Be star companion at periastron. These outbursts last
for a few days to tens of days and are different from the Type~II outbursts
which are caused by enhanced episodic outflow of the Be Star. Type~II
outbursts, are infrequent giant outbursts with peak luminosities reaching as
high as $L_{\rm X} = 10^{38}$ erg~s$^{-1}$ or more. These outbursts last for
several weeks to months. The timing of these outbursts is not related to any
underlying orbital period of the system. These Type~II outbursts occur when a
large fraction of the Be star's disk is believed to be accreted. The Type~I
outbursts, though they occur only around the time of the periastron passages,
do not occur in all the binary orbits and often a series of Type~I outbursts
are seen following a large Type~II outburst. Fig.~\ref{X-ray_outburst} shows
the presence of Type~I and Type~II X-ray outbursts in the \textit{Swift}/BAT
light curves (in 15--50 keV energy range) of A0535$+$262 and EXO~2030$+$375
Be/X-ray binary systems.

Supergiant Fast X-ray Transients (SFXTs) is a new class of high mass X-ray
binaries, discovered with \textit{Integral} observations (Sguera et al.\ 2005),
that are associated with OB supergiant stars. These objects are characterized
by the occurrence of very fast X-ray outbursts. Outside the outbursts, they are
not detected as X-ray sources or only detected as a weak source. In quiescence,
the X-ray luminosity of these objects is found to be $\sim 10^{32}$
erg~s$^{-1}$ (Bozzo et al.\ 2010), with a dynamic range of 3--5 orders of
magnitude for the outbursts. The X-ray outbursts in these systems are
characterized by bright flares with a duration of a few hours (lasting from
$\sim 3$ to $\sim 8$ hours, rarely even a few days), very sharp rise, reaching
the peak of the flare in $\leq 1$ h. Since the SFXTs are very bright in X-ray
band but for very short duration (during the short transient outburst), these
objects are difficult to detect with the X-ray sky monitors. During the
outburst, the X-ray spectra of these sources are found to be hard and similar
to that of other HMXBs hosting accreting neutron stars. Even though pulse
periods have only been measured for a few SFXTs e.g.\ IGR~J11215$-$5952 ($\sim
187$ s; Swank et al.\ 2007) and IGR~J18410$-$0535 ($\sim 4.7$ s; Sidoli et al.\
2008), it is likely that all SFXTs might host a neutron star. The
mechanism producing these short outbursts is still being debated. It is
probably related to either the properties of the wind from the supergiant
companion (Negueruela et al.\ 2008 and references therein) or to the presence
of a centrifugal or magnetic barrier (Grebenev \& Sunyaev 2007; Bozzo, Falanga
\& Stella 2008).

Here we discuss various important temporal, spectral, and joint
temporal-spectral properties of HMXBs in which there have been significant
recent developments. Then we discuss some other emerging important issues of
HMXBs, like the SFXT phenomena. We also briefly discuss the optical/IR
properties of the transient HMXBs during the outbursts and mention the
potential of the upcoming mission Astrosat in furthering investigations of this
important class of sources.

\section{Aspects of transient HMXBs probed during the outbursts}

The transient HMXBs show change of X-ray intensity by several orders of
magnitudes during the outbursts, most likely due to changes in the mass
accretion rate. This allows probe of several features like the pulse shape,
accretion torque, quasi-periodic oscillations, energy spectrum, cyclotron
absorption features in the spectrum over a large range of X-ray
luminosity. Changes in the mass accretion rate is expected to cause changes in
the accretion process in a complex way, for example by changes in the inner
disk radius, accretion torque, height of the accretion column, X-ray beaming
pattern, and even temperature of the electrons that produce the hard X-rays by
Compton up-scattering. In the last few years, availability of several medium
and hard X-ray transient monitors, several narrow field X-ray instruments with
excellent timing and spectral capabilities and wide energy coverage, and
intense observation programs have enabled very detailed investigations of
transient HMXBs. Some of the important aspects are discussed below.

\subsection{Temporal properties}

Observations with the Proportional Counter Array (PCA) instrument of the 
Rossi X-ray Timing Explorer (\textit{RXTE})
has been most important in investigation of the temporal studies of all kinds
of X-ray sources, including the transient HMXBs. In addition, the 
Gamma-ray Burst Monitor (GBM)
instrument onboard Fermi has also been a major contributor in studies of pulsar
spin-up and QPOs. The temporal characteristics of the SFXTs have primarily
been revealed from \textit{Swift} and \textit{Integral} observations.

\subsubsection{Luminosity dependence of pulse profiles}

Accretion-powered X-ray pulsars, both transient and persistent sources, are
known to show luminosity dependence of the pulse profile. Differences in shape
of X-ray pulse profiles are considered to be due to the differences in the
geometrical configuration of the accretion column around the magnetic axis
(Nagase 1989). The structure of the  accretion column near the magnetic poles
of the neutron star determines the basic profile of the pulse pattern. A most
remarkable example of pulse profile evolution was investigated in
EXO~2030$+$375 (Parmar, White \& Stella 1989). The evolution of the pulse
profile during the outbursts indicates changes in the accretion flow from the
inner accretion disc to the neutron star. Along with the luminosity dependence,
most of the accretion-powered pulsars also show strong energy dependence of the
pulse profile, discussed in Section 4.

Luminosity dependence of pulse profiles are seen in several transient HMXB
pulsars. In one of the nearby and brightest Be/X-ray binary pulsar A0535$+$262,
the pulse profile is known to be single peaked during the quiescent phase
(Mukherjee \& Paul 2005) whereas it is double-peaked and complex during the
X-ray outbursts (Naik et al.\ 2008; Caballero et al.\ 2007). An extensive study
of the pulse profiles of the recent outbursts of two transient pulsars
GX~304$-$1 and 1A~1118$-$61 showed significant evolution during the outburst
(Devasia et al. 2011a, 2011b; Maitra, Paul \& Naik 2011). Apart from the energy
and luminosity dependence of pulse profiles, dips or dip-like features in the
pulse profiles are also seen in many transient X-ray pulsars. The presence of
dips/dip-like structures in the pulse profiles of these X-ray pulsars is
described as due to the obscuration of X-ray radiation from the region near the
magnetic poles by dense matter in the accretion streams along which the matter
flows from the inner accretion disk to the neutron star. The associated
spectral signatures of this are discussed in Section 4.

\subsubsection{Accretion torque}

In accretion powered binary X-ray pulsars, the flow of material from the
companion star to the neutron star is interrupted when the magnetic stress
starts dominating over the material stresses at the magnetospheric radius
$r_{\rm m}$. Matter becomes attached to the magnetic field lines at $r_{\rm m}$
and gets transported to the magnetic poles of the neutron star. The torque of
the infalling material captured from the accretion disk and the torque
transferred by the accretion disk to the neutron star through the magnetic
field lines cause changes in the neutron star spin-rate. At high mass accretion
rate, during the outbursts, the neutron stars experience an accretion torque
which leads to spinning up of the neutron star. At low accretion rate, like
during the quiescent period between the outbursts, the neutron stars often
experience net negative torque (Bildsten et al.\ 1997). At very low accretion
rates, if the magnetospheric radius lies outside the corotation radius $r_{\rm
co}$, where the Keplerian orbital frequency equals the spin frequency of the
neutron star, matter may be expelled from the system. In such cases, accretion
can be centrifugally inhibited which is called ``propeller effect''. During
this regime, the neutron star spins down rapidly and pulsations may not even be
detectable. In several cases, disappearance of pulses at low flux rate have
been shown as evidence of the propeller effect (Cui 1997 and references
therein). However, there are also counter examples (Mukherjee \& Paul 2005;
Naik, Paul \& Callanan 2005). Due to this complex nature of the relation
between the mass accretion and net torque onto the neutron star, the transient
sources are very good candidates to study the interaction between material in
the accretion disk and the magnetic field through measurements of rate of
change of spin period and X-ray luminosity.

The accretion torque theory predicts that the magnetospheric radius should
decrease with increase in the rate of mass accretion as per the relation
$r_{\rm m} \propto \dot{M}^{-2/7}$ for $r_{\rm m} < r_{\rm co}$ (Ghosh \& Lamb
1979). This implies that a neutron star spins up with a rate related to the
mass accretion rate as $\dot{\nu} \propto \dot{M}^{-6/7}$. Such a dependence
was tested in several transient X-ray pulsars. A correlation between the
spin-up rate and X-ray luminosity has been observed during outburst of several
transient binary X-ray pulsars such as EXO~2030$+$375 (Reynolds et al.\ 1996,
Parmar et al.\ 1989), 2S~1417$-$62 (Finger, Wilson \& Chakrabarty 1996),
A~0535$+$262 (Bildstein et al. 1997), GS~0834$-$43 (Wilson et al.\ 1997),
GRO~J1744$-$28, (a LMXB pulsar; Bildsten et al.\ 1997). Very recently, the
long-term spin frequency evolution history of A0535$+$262 (since 1975 to 2006)
was analyzed in detail. It was found that this Be/X-ray pulsar shows a global
spin-up trend in which short, rapid spin-up episodes during the outbursts are
followed by extended, slow spin-down during the quiescent periods
(Camero-Arranz et al.\ 2011). From the long term spin frequency evolution study
of this pulsar with the \textit{Fermi/GBM} a strong correlation was found between the
pulse flux and the spin-up rate which is in agreement with the earlier findings
in A0535$+$262 and other transient pulsars.

\subsubsection{Quasi-periodic oscillations and its relation with X-ray
luminosity}

Quasi-periodic oscillations (QPOs) in X-ray binary pulsars are thought to be
related to the motion of inhomogeneously distributed matter (blobs) in the
inner accretion disk, and they provide useful information about the interaction
between accretion disk and the neutron star. In X-ray pulsars, the QPO
frequency ranges from $\sim 1$ mHz to $\sim 40$ Hz, and they can be from $\sim
100$ times smaller to $\sim 100$ times larger than the pulsar spin frequencies
(Psaltis 2006). The presence of QPOs in the power density spectrum of X-ray
pulsars is generally explained with the Keplerian frequency model (van der Klis
et al.\ 1987) or the magnetospheric beat frequency model (Alpar \& Shaham
1985). In Keplerian frequency model, the QPOs arise from the modulation of the
X-rays by inhomogenities in the accretion disc, at the Keplerian frequency. In
this model, the QPO frequency is same as that of the Keplerian frequency of the
inner accretion disk. When the Keplerian frequency at the inner edge of the
accretion disk is below the neutron star spin frequency, centrifugal forces are
expected to inhibit accretion. Keplerian frequency model, therefore, can be
applicable only when the QPO frequency is above the neutron star spin
frequency, as seen in EXO~2030$+$375 (Angelini, Stella \& Parmar 1989), A0535$+$262
(Finger et al.\ 1996). In the beat frequency model, blobs of matter orbits
the neutron star approximately at the Keplerian frequency of the inner edge of
the accretion disk, accreting at a rate that is modulated by the rotating
magnetic field. This produces power spectral feature at the beat frequency
between Keplerian frequency and spin frequency of the neutron star. According
to this model, the QPO frequency is equal to the difference between the
Keplerian frequency of the inner accretion disk and the spin frequency of the
pulsar.

\begin{figure}
\centerline{\includegraphics[angle=-90,width=6cm]{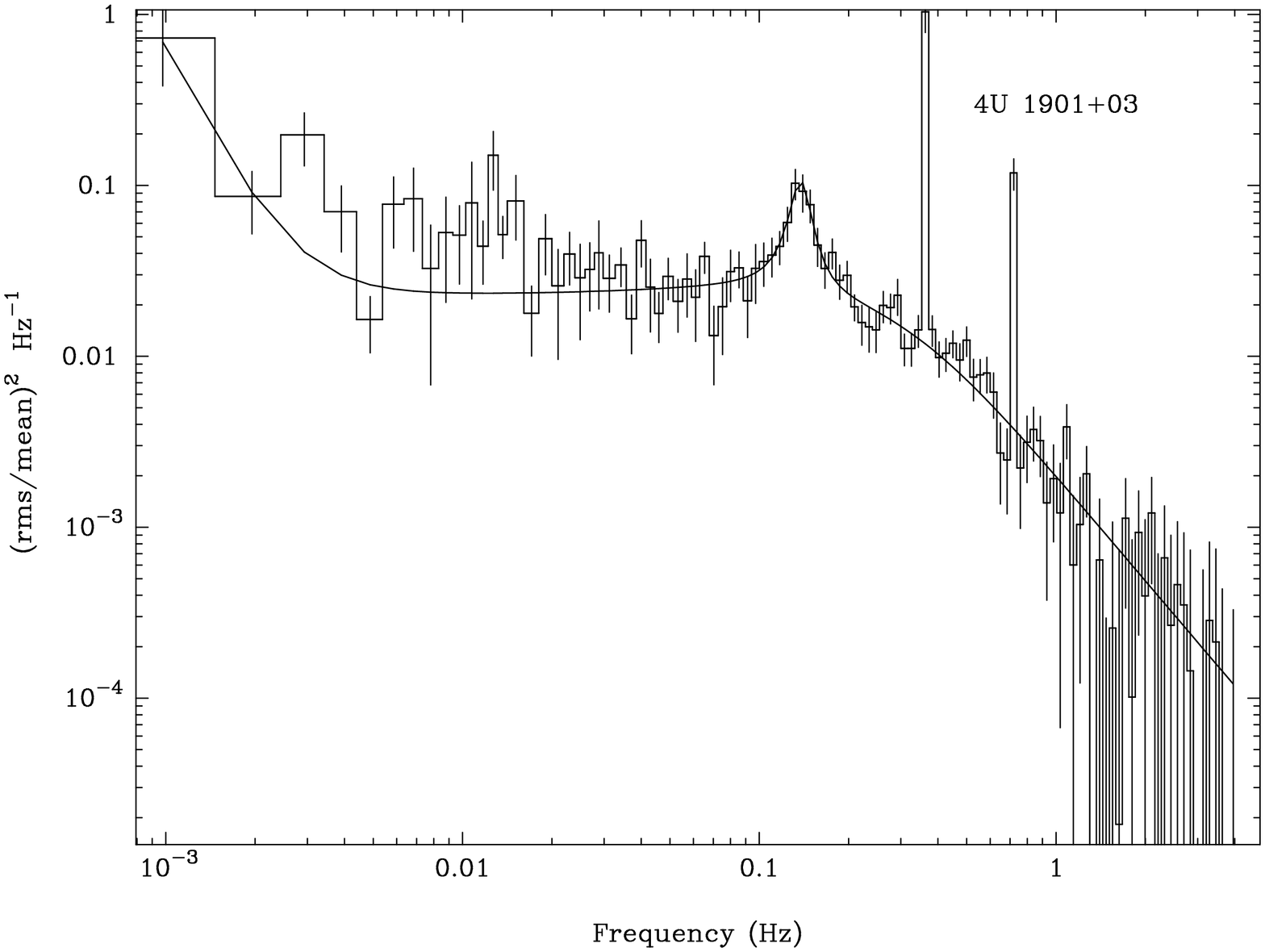} \qquad
            \includegraphics[angle=-90,width=6cm]{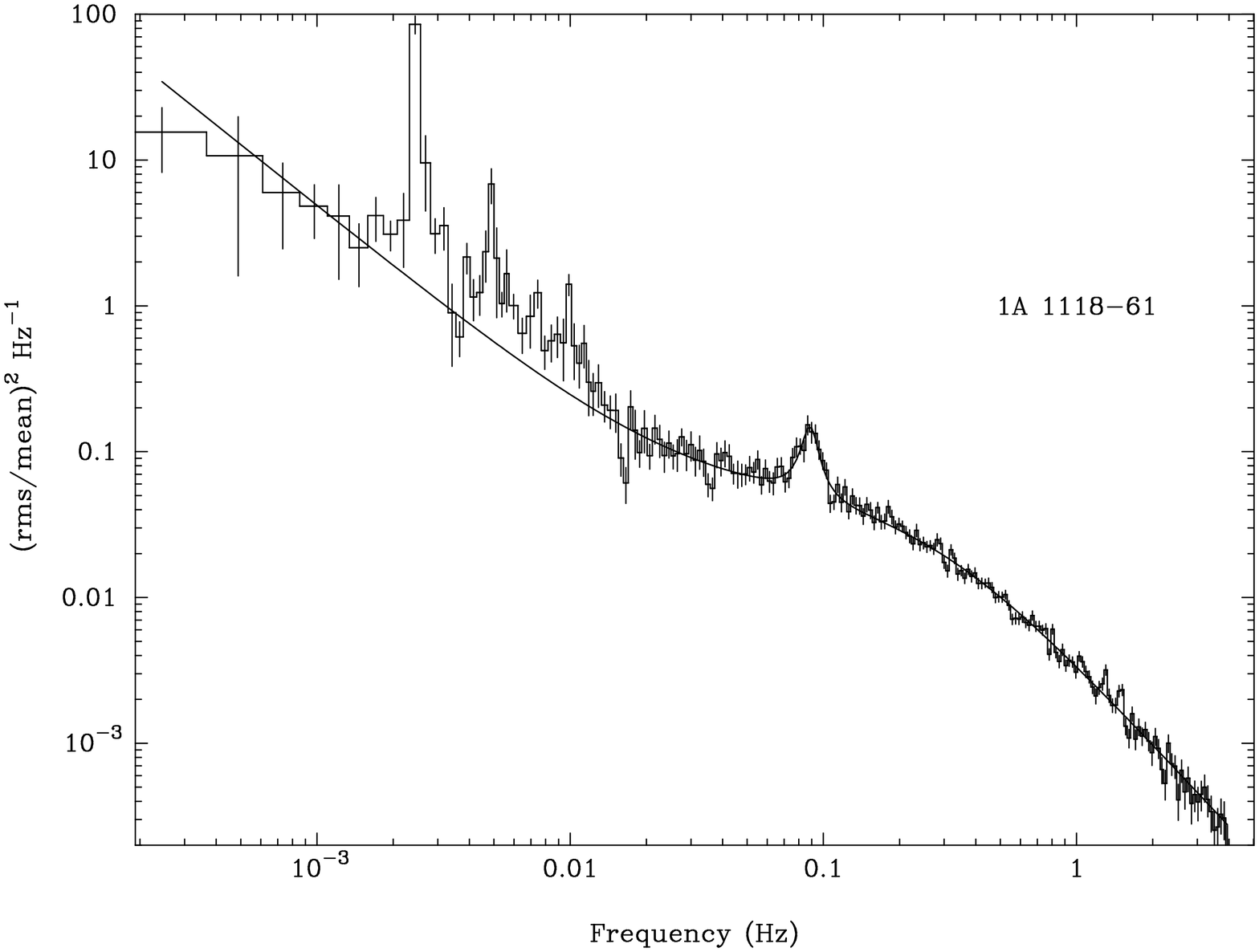}}
\medskip
\centerline{\includegraphics[angle=-90,width=6cm]{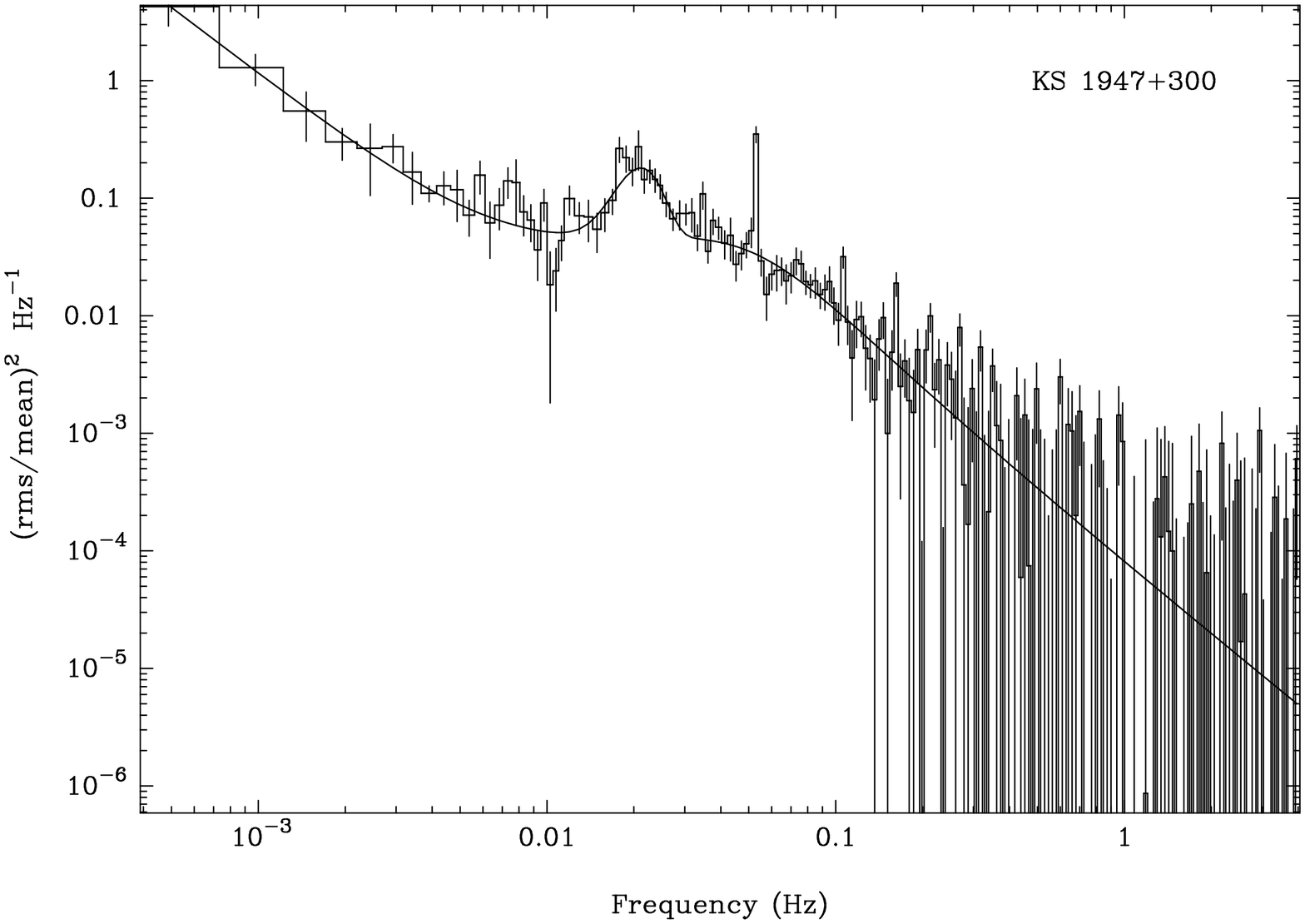} \qquad
            \includegraphics[angle=-90,width=6cm]{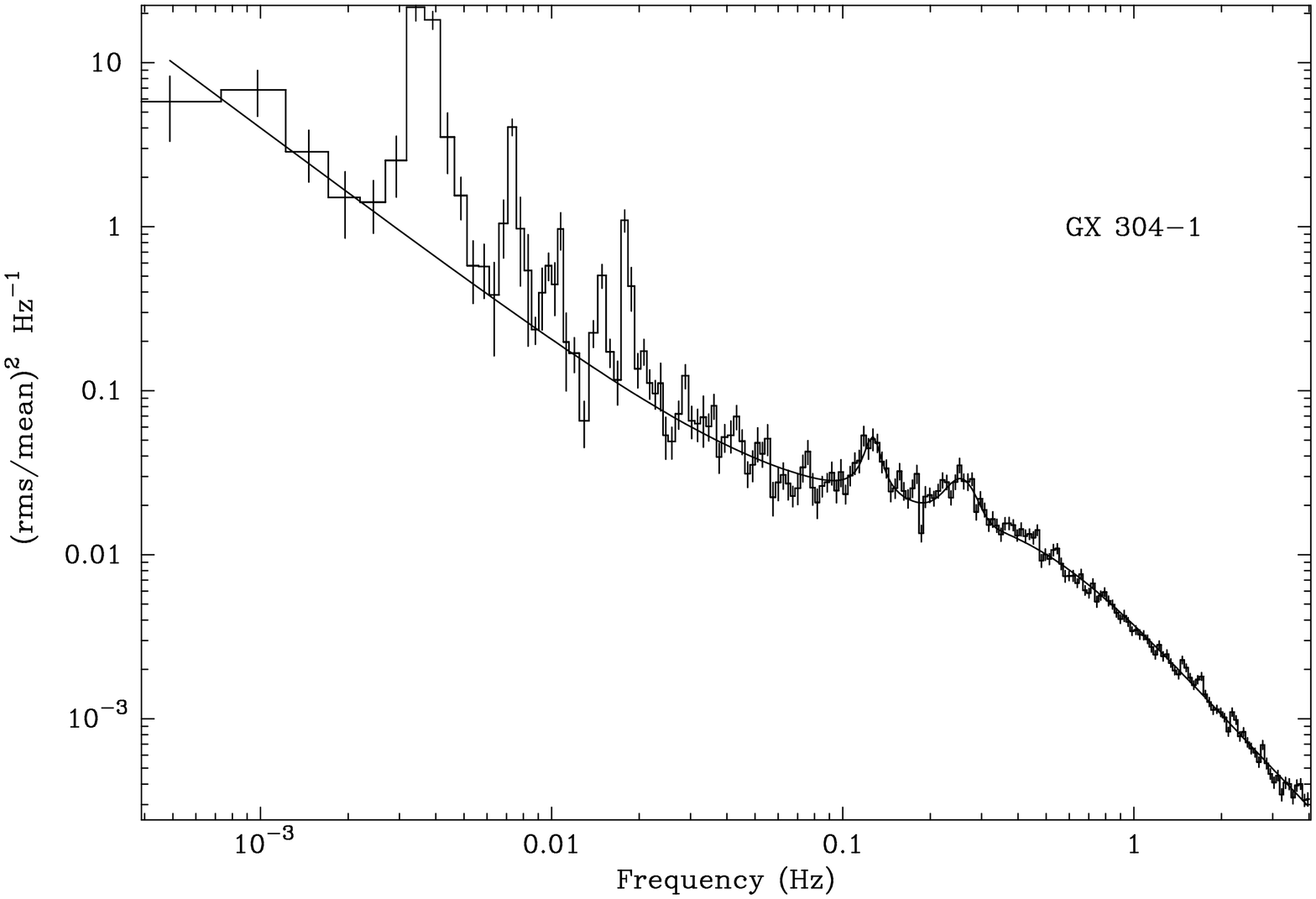}}
\caption{The quasi periodic oscillations discovered recently in four transient
accretion powered X-ray pulsars are shown here. ({Top:} 4U~1901$+$03 and
1A~1118$-$61; {bottom:} KS~1947$+$300 and GX 304$-$1). The pulse frequency
peaks and its harmonics are also seen along with the broad QPO
features.}\label{qpo}
\end{figure}

In accretion powered X-ray pulsars, QPOs seem to occur more in transient
sources compared to the persistent ones. Transient HMXB pulsars from which QPOs
have been detected are KS~1947$+$300 (20 mHz; James et al.\ 2010), SAX
J2103.5$+$4545 (44 mHz; Inam et al.\ 2004), A0535$+$262 (50 mHz; Finger et al.
1996), V0332$+$53 (51 mHz; Takeshima et al.\ 1994), and 4U~0115$+$63 (62 mHz;
Soong \& Swank 1989), XTE~J1858$+$034 (110 mHz; Paul \& Rao 1998; Mukherjee et
al. 2006), EXO~2030$+$375 (200 mHz; Angelini et al.\ 1989), XTE J0111.2$-$7317
(1270 mHz; Kaur et al.\ 2007), 4U 1901$+$03 (James et al.\ 2011), 1A~1118$-$61
(Devasia et al. 2011a), MAXI J1409$-$619 (Kaur et al.\ 2010) and GX~301-4
(Devasia et al. 2011b). The power density spectra of four sources are shown in
Fig.~\ref{qpo}, in which the QPO features are clearly seen. In two of the
sources the QPO frequency is higher than the spin frequency and it is opposite
for the remaining two sources. It has been found that the QPOs are rare and
transient events in HMXB pulsars. In transient HMXB pulsars which show QPOs,
the feature is not detected always during all the observed X-ray outbursts. In
some sources (e.g.\ A0535$+$262), the QPO feature is present all along in some
of the outbursts while completely absent in some other outbursts (Finger et al.
1996; Camero-Arranz et al.\ 2011). While in some other sources (4U~1901$+$03
and 1A~1118$-$61), the QPOs are detected only near the end of the outburst,
when the X-ray intensity has decreased to a small fraction of the peak
intensity. At high mass-accretion rate (during X-ray outbursts), the accretion
disc is expected to extend closer to the neutron star. A positive correlation
between the QPO centroid frequency and the X-ray intensity is, therefore,
expected in the transient HMXB pulsars (Finger 1998). A strong correlation
between the QPO centroid frequency with the X-ray intensity and neutron star
spin-up rate was seen in the transient Be/X-ray binary pulsar A0535$+$262
(Finger et al.\ 1996; Camero-Arranz et al.\ 2011). Similar correlation between
the QPO frequency and the source intensity was found in case of the transient
HMXB pulsar EXO~2030$+$375 (Angelini et al.\ 1989). In case of XTE~J1858$+$034,
though the correlation between QPO frequency and the X-ray flux was not clearly
seen, the QPO frequency and the one day averaged X-ray flux decreased with time
during the April--May 2004 X-ray outburst (Mukherjee et al.\ 2006; Paul \& Rao
1998). This suggests that the observed QPO frequency, like the spin-up rate and
X-ray flux, is controlled by the mass accretion rate.

\subsubsection{Orbital parameters, orbital evolution}

In the X-ray binaries with eccentric orbits, tidal interactions may result in
circularisation and apsidal motion (Lecar, Wheeler \& McKee 1976) that can be 
measured if
the compact object is an X-ray pulsar. Apsidal motion is the rate of change of
the longitude of periastron of the binary orbit. The rate of apsidal motion is
directly related to the distribution of mass in the tidally distorted companion
star and hence is directly related to the stellar structure constant of the
companion star. The transient Be X-ray binaries, which often have accretion
powered X-ray pulsars are suitable candidates to measure the orbital parameters
and measurements carried out during multiple outbursts over several years can
be used to estimate the apsidal motion constant. In addition to the tidal
effect, the orbits of the X-ray binary systems evolve due to mass transfer from
the companion star to the compact object, mass loss from the binary system in
the form of stellar wind from the companion star, and/or gravitational wave
radiation (Verbunt 1993). In the case of HMXB systems which often have orbital
period between a few days to a few hundred days, however, the effect of
gravitational wave radiation and even that of mass transfer is weak compared to
the other processes. The best way to measure orbital evolution of X-ray
binaries is by repeated measurements of orbital parameters by pulse timing.
However, measurement of orbital parameters of the transient HMXBs by pulse
timing requires long X-ray observations during the outbursts. The complete set
of orbital parameters is known only in 18 Be/X-ray binary systems out of a
total of $\sim$130 sources in the catalogue of Be/X-ray binaries in our Galaxy,
the Large Magellanic Cloud (LMC) and the Small Magellanic Cloud (SMC) (Raguzova
\& Popov 2005).

Using observations made with the \textit{RXTE}-PCA, the complete set of orbital
parameters have recently been measured for the transient Be/X-ray binary
pulsars V0332$+$53 and 2S~1417$-$624 along with a successful measurement of the
rate of apsidal motion in 4U~0115$+$63 (Raichur \& Paul 2010). For the HMXB
pulsar 4U~0115$+$63, the \textit{RXTE} observations of the X-ray outbursts in 1999 and
2004 were used to determine the changing orbital parameters, and along with the
previous measurements the rate of apsidal motion was determined to be
($\dot\omega = 0^\circ.04 \pm 0^\circ.02$ yr$^{-1}$). Orbital parameters of
several other HMXB transients have been measured accurately in the Milky Way.
Similar measurements are also being done in case Be/X-ray binaries in the Small
Magellanic Could (SMC) using data from the \textit{RXTE} observatory. The
Be/X-ray binary systems in the low-metallicity environment in the SMC show
similar behaviour in the distribution of orbital periods and eccentricities as
that of the HMXBs in our Galaxy. This suggests that metallicity may not play an
important role in the evolution of such systems (Townsend et al.\ 2011).

\subsubsection{Power Density Spectra of HMXBs}

Most of the accretion powered X-ray pulsars, in addition to their periodic
modulations, show \textit{aperiodic} X-ray variabilities in a wide range of
time scales. These variabilities in the X-ray intensity are reflected as
various different types of features in the Power Density Spectra (PDS).
Aperiodic variability is more common than periodic variability in X-ray binary
systems as the accretion flow from the companion to compact object is generally
turbulent. Characterizing the aperiodic variabilities of X-ray sources is an
important step in understanding the complicated dynamics of the accretion disk
around the compact objects. The PDS of accreting X-ray pulsars where the
accretion disk at the magnetospheric boundary could be close to being in
corotation with the neutron star, have a distinct break/cut-off around the
neutron star spin frequency (Hoshino \& Takeshima 1993). In these systems, the
break in the PDS probably represents the transition from the disk-like
accretion flow to magnetospheric flow at the frequency characteristic of the
magnetospheric radius.

\section{X-ray spectrum and spectral evolution during the outburst}

It has already been discussed above that the majority of the HMXBs are known to
be Be/X-ray binaries. Mass transfer from the Be companion to the neutron star
takes place through the circumstellar disk. Strong X-ray outbursts are normally
seen when the neutron star passes through the circumstellar disk, often during
the periastron passage. Transient X-ray pulsars in general are prone to wide
variations in mass accretion rates and are suitable systems to test the various
accretion regimes onto high magnetic field neutron stars. The X-ray luminosity
of these transient X-ray pulsars is highly variable i.e.\ from 10$^{32}$ --
10$^{33}$ erg s$^{-1}$ during quiescence to 10$^{36}$ -- 10$^{37}$ erg s$^{-1}$
during the peak of the outbursts. Though the mass accretion onto neutron stars
in binary systems is a complex physical process, broad band X-ray spectra of
the X-ray pulsars can be described by relatively simple models. A considerable
fraction of the photons emitted from the neutron star gets reprocessed by the
matter along the line of sight, accretion column, accretion disk before
reaching the observer. Around the periastron passage, there is possibility of
an increase in the column density of matter along the line of sight. Close to
the neutron star, the accreting matter is phase locked in the strong magnetic
field and these accretion streams can also cause absorption and reprocessing of
the X-rays.

The spectra of the binary X-ray pulsars are generally described by a power-law,
broken power-law or power-law with high energy cutoff continuum models. In some
cases, the pulsar spectrum has also been described by the Negative and Positive
power-law with EXponential cut-off (NPEX) continuum model which is an
approximation of the unsaturated thermal Comptonization in hot plasma
(Makishima et al.\ 1999). This continuum model reduces to a simple power-law
with negative slope at low energies that is used to describe the spectra of
accretion powered X-ray pulsars at low energies. The analytical form of the
NPEX model is
\begin{eqnarray} \nonumber
  NPEX(E) = (N_1 E^{-\alpha_1} + N_2 E^{+\alpha_2})
            \exp \left( -\frac{E}{kT} \right)
  \label{eq1}
\end{eqnarray}
where $E$ is the X-ray energy (in keV), $N_1$ and $\alpha_1$ are the
normalization and photon index of the negative power-law respectively, $N_2$
and $\alpha_2$ are those of the positive power-law, and $kT$ is the cutoff
energy in units of keV. The first power-law is the usual high energy component
produced by inverse Compton scattering of the soft X-ray photons and the second
power-law describes the black body radiation, if the corresponding photon index
$\alpha_2$ is close to 2. The physical motivation for this model is the
coupling between the accretion hot spot and the up-scattered X-ray photons in
the accretion column, which results in a cutoff power-law.

In case of a few X-ray pulsars, it has been reported that the absorption has
two different components (Endo, Nagase \& Mihara 2000; Mukherjee \& Paul 2004). In this
model, one absorption component absorbs the entire spectrum whereas the other
component absorbs the spectrum partially. This model is known as partial
absorption model. The partial covering model can also be described as
consisting of two power-law continua with a common photon index but with
different absorbing hydrogen column densities. The partial covering absorption
model is applicable if the second absorbing component is smaller in size
compared to the emission region. The analytical form of the partially covering
high energy cutoff power-law model is
\begin{eqnarray}\nonumber
   N(E) = {e^{-\sigma(E)N_{\rm H1}}}
          \left( S_{1}+S_{2}e^{-\sigma(E)N_{\rm H2}} \right)
          {E^{-\Gamma}}{I(E)}
\end{eqnarray}
where
\begin{eqnarray}\nonumber
  I(E) = \cases{
           1 & \textrm{for $E < E_{\rm c}$,} \cr
           e^{ -\left( {E-E_{\rm c} \over E_{\rm f}} \right) }
             & \textrm{for $E > E_{\rm c}$,}
         }
\end{eqnarray}
$N(E)$ is the measured spectrum, $\Gamma$ is the photon index, $N_{\rm
H1}$ and
$N_{\rm H2}$ are the two equivalent hydrogen column densities, $\sigma$ is
the photo-electric cross-section, $S_{1}$ and $S_{2}$ are the normalizations of
the two power law components, $E_{\rm c}$ is the cut-off energy and
$E_{\rm f}$ the $e$-folding energy. According to this model, a part of the
continuum source is obscured, resulting in a harder spectrum. If the absorbing
component is in the form of an accretion stream or is a part of the accretion
column, it can be phase locked with the neutron star, resulting into a
phase-dependent column density and covering fraction. At higher energy,
however, the partial covering absorption model is the same as a simple
power-law model. Some results from pulse phase resolved spectroscopy of
transient pulsars is further discussed in Section 4. A narrow iron fluorescence
line at 6.4 keV is often found in the spectrum of HMXB pulsars.

\begin{figure}
\centering
\includegraphics[width=3.0 cm, angle=-90]{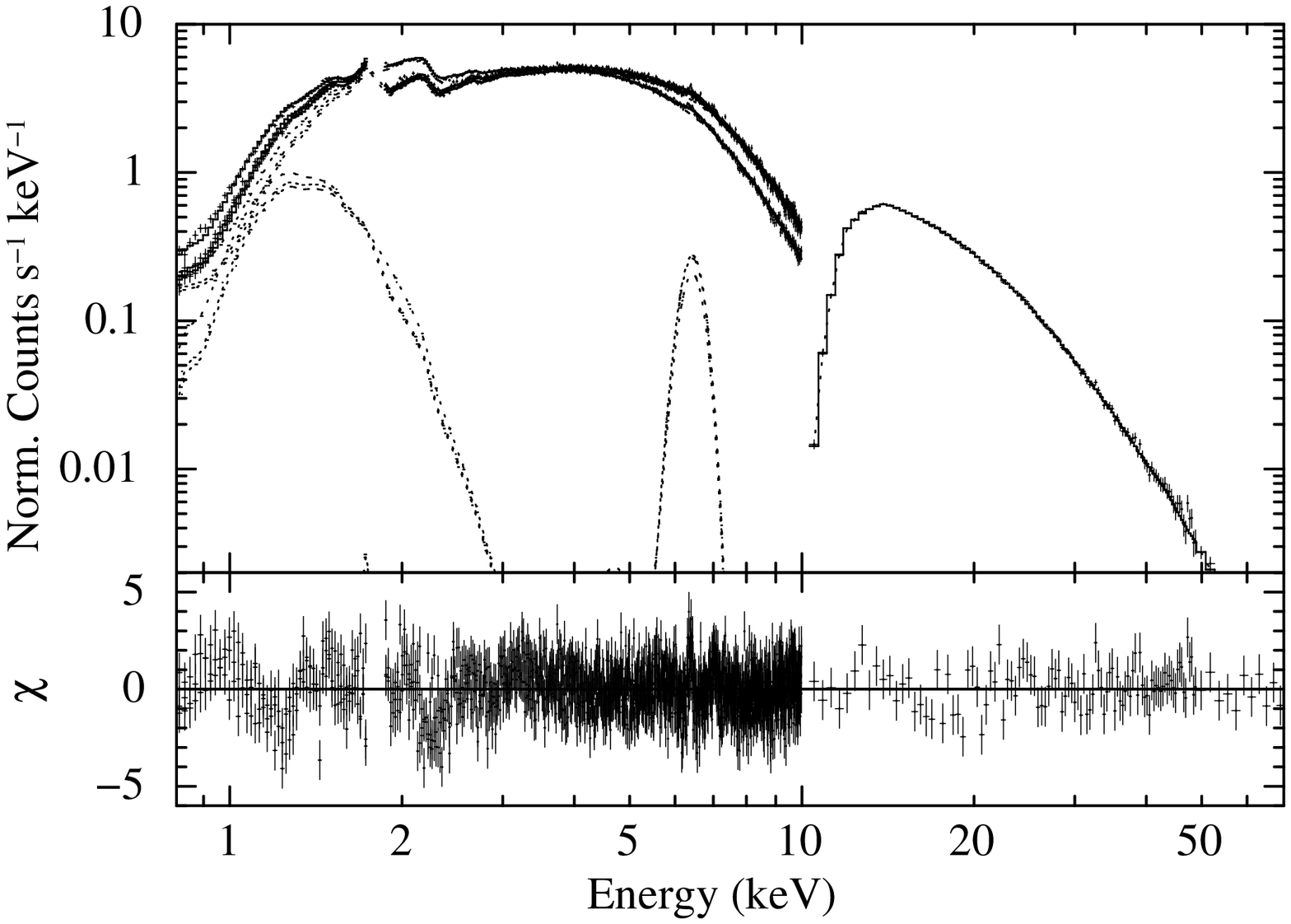}
\includegraphics[width=3.0 cm, angle=-90]{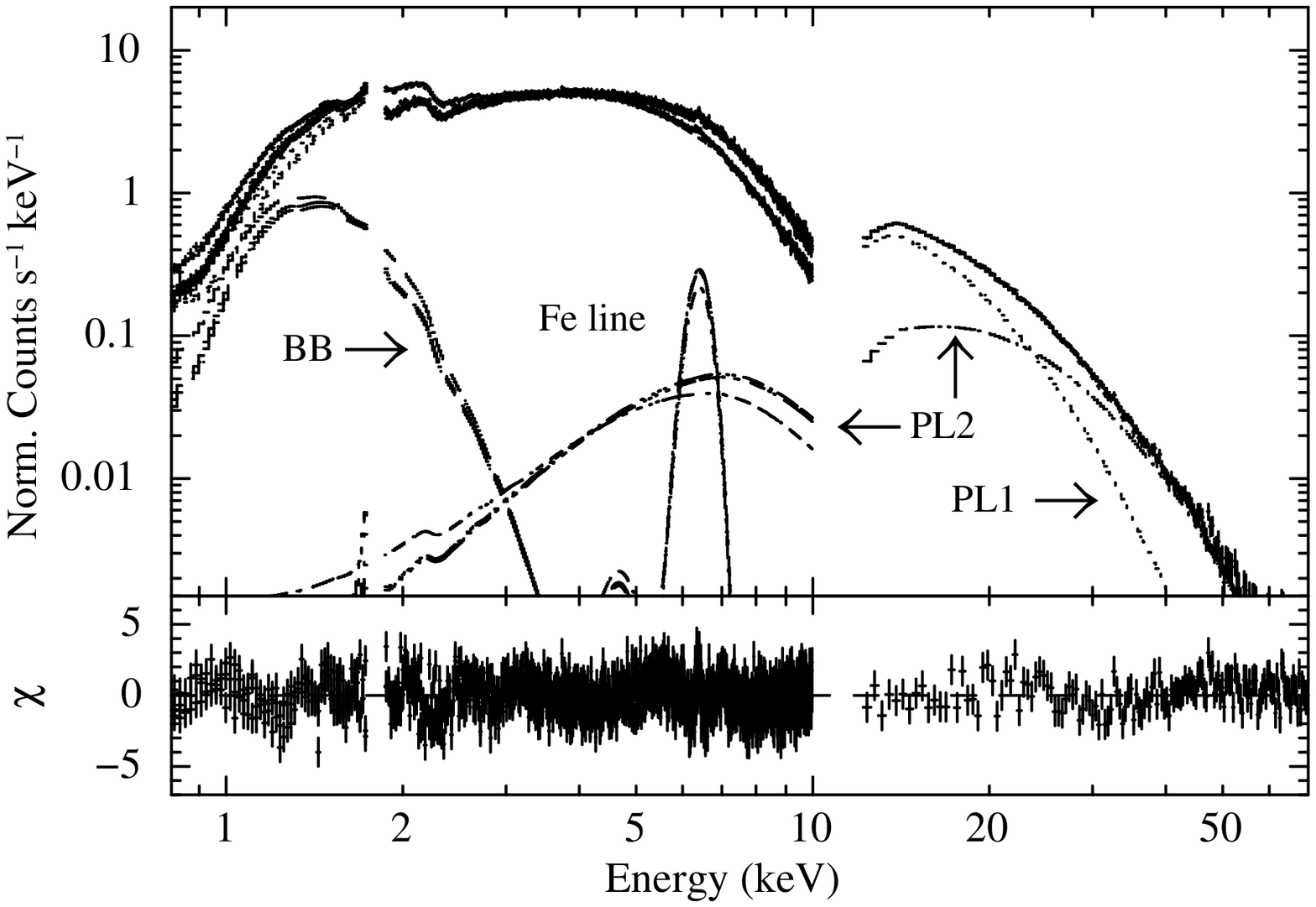}
\includegraphics[width=3.0 cm, angle=-90]{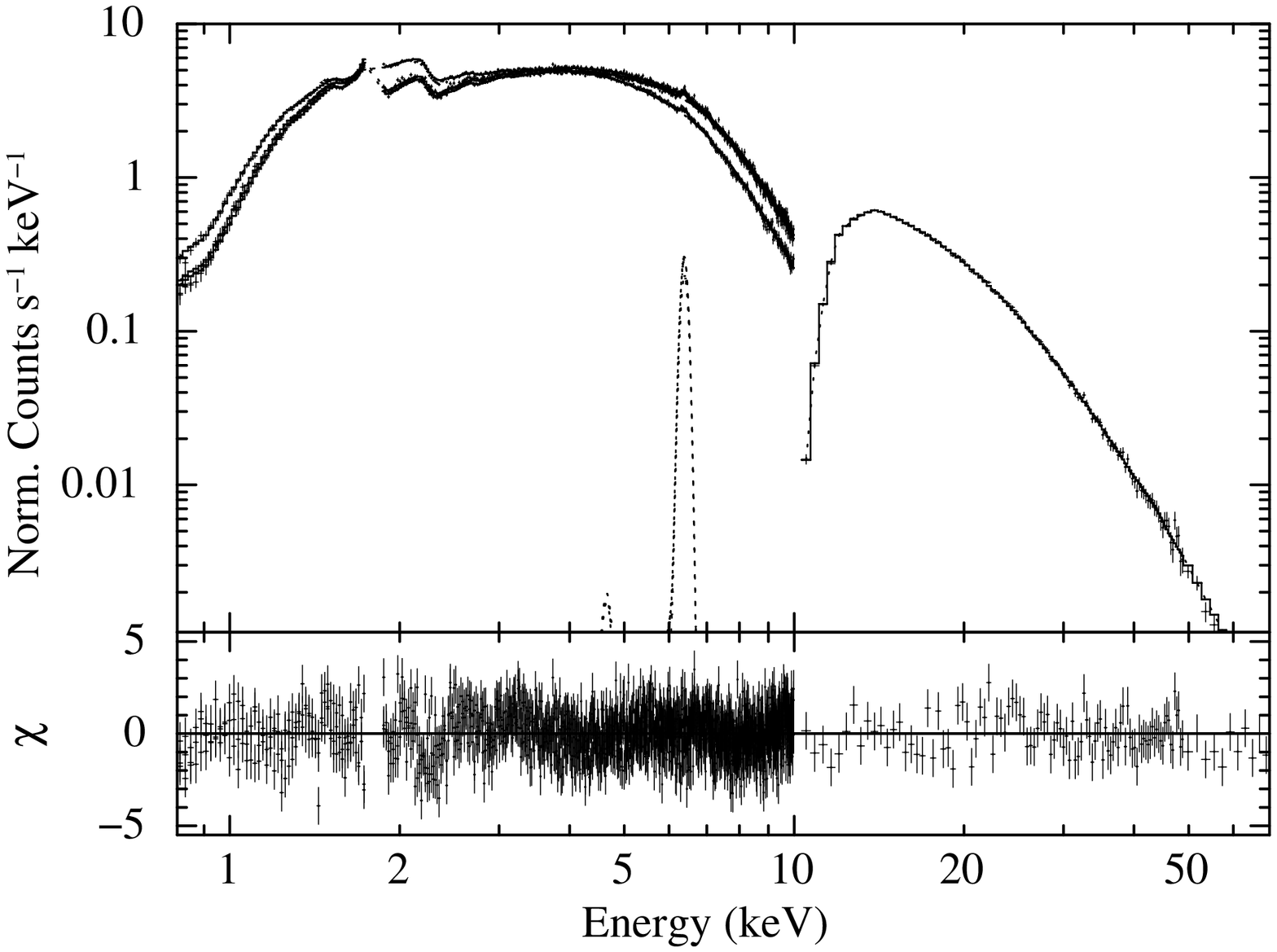}
\caption{Energy spectrum of GRO~J1008$-$57 obtained with the XIS and PIN
detectors of \textit{Suzaku} observation, along with the best-fitting models
comprising (a) a blackbody component, high energy cut-off power-law continuum
model and a narrow iron line emission (left panel), (b) a blackbody (BB)
component, NPEX continuum model and a narrow iron (Fe line) emission (middle
panel), and (c) a partially absorbed high energy cut-off power-law continuum
model and a narrow iron line emission (right panel). The bottom panels show the
contributions of the residuals to the $\chi^2$ for each energy bin for the
best-fitting models.} \label{gro_spec} \end{figure}

Broad-band spectrum of the Be/X-ray binary pulsar GRO~J1008$-$57 was found to
be well fitted with all three continuum models as described above. The count
rate spectra of GRO~J1008$-$57 are shown in Fig.~\ref{gro_spec} along with
the model components (top panels) and residuals to the best-fitting model
(bottom panels). In case of the hard X-ray transient pulsar 1A~1118$-$61, 3--30
keV energy spectra obtained from several \textit{RXTE} observations during an X-ray
outburst in 2009 were found to be well described with a partial covering
power-law model with a high-energy cut-off and an iron fluorescence line
emission (Devasia et al.\ 2011a). Broad-band spectroscopy of \textit{Suzaku}
observations of the pulsar also revealed that the partial covering power-law
model is the suitable model to describe the spectra (Maitra et al.\ 2011). The
spectrum of transient HMXB pulsar GX~304$-$1 was found to be better described
by the partial covering power law with high energy cut-off model than the power
law with high energy cut-off model (Devasia et al.\ 2011b). The partial
covering absorption model is found to be most suitable continuum model even in
case of persistent HMXB pulsars such as GX~301$-$2 (Mukherjee \& Paul 2004) and
Cen~X-3 (Naik, Paul \& Ali 2011).

\begin{figure}
\centering
\includegraphics[height=3.1in, width=3.5in, angle=-90]{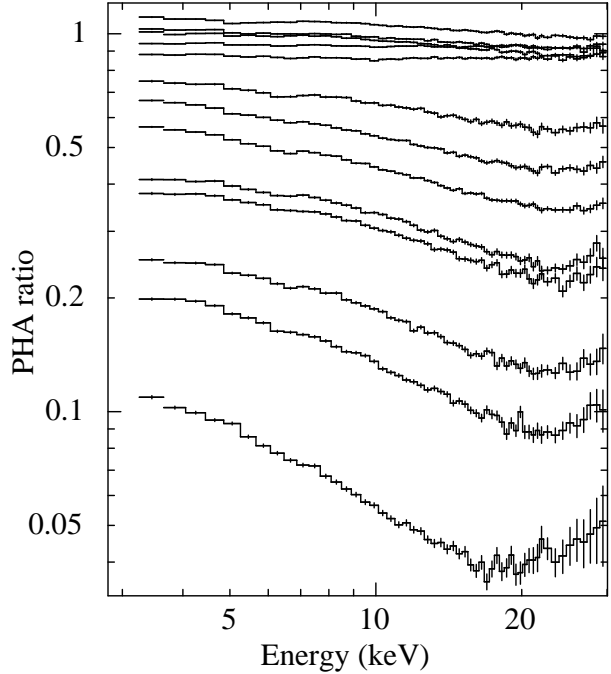}
\caption{Ratios of the 3--30 keV X-ray spectrum obtained on different days of
{RXTE} observations of GX~304-1 during its 2010 August outburst with the
spectrum obtained on August 15 near the peak of the outburst are shown. A
softening of the spectrum below about 18 keV and a hardening above 18 keV is
evident during the decay of the outburst. The figure is taken from Devasia et
al. (2011b).} \label{gx304_sr} \end{figure}

In order to investigate the spectral evolution of the transient HMXB pulsars
during outbursts, frequent X-ray observations of the pulsar are required.
Spectral evolution during the outbursts has been detected in several transient
sources, in the simple form of change in hardness ratio (Reig 2008). Detailed
study of spectral evolution during the outbursts have recently been carried for
a few sources. Extensive \textit{RXTE} observations of the transient HMXB pulsar
GX~304-1 during an outburst in 2010 August was reported by Devasia et al.
(2011b). They found that the 3--30 keV spectrum was well fitted with a partial
covering power law model with a high energy cut-off and iron fluorescent line
emission. Significant spectral evolution was seen during the outburst. The
ratio of the spectra during the entire outburst to the spectrum on one epoch
(close to the peak of the outburst; as shown in Fig.~\ref{gx304_sr}), showed
the softening of the spectrum below $\sim 18$ keV and hardening beyond $\sim
18$ keV, during the decay of the outburst. Similar spectral evolution during
the transient outbursts in HMXB pulsars has also been studied in 1A~1118$-$61
(Devasia et al.\ 2011a). The spectral changes indicate corresponding changes in
the emission region with variations in the mass accretion rate.

\subsection{Cyclotron lines in transient HMXB pulsars}

\begin{figure}
\centering
\includegraphics[height=3.0in, width=2.3in, angle=-90]{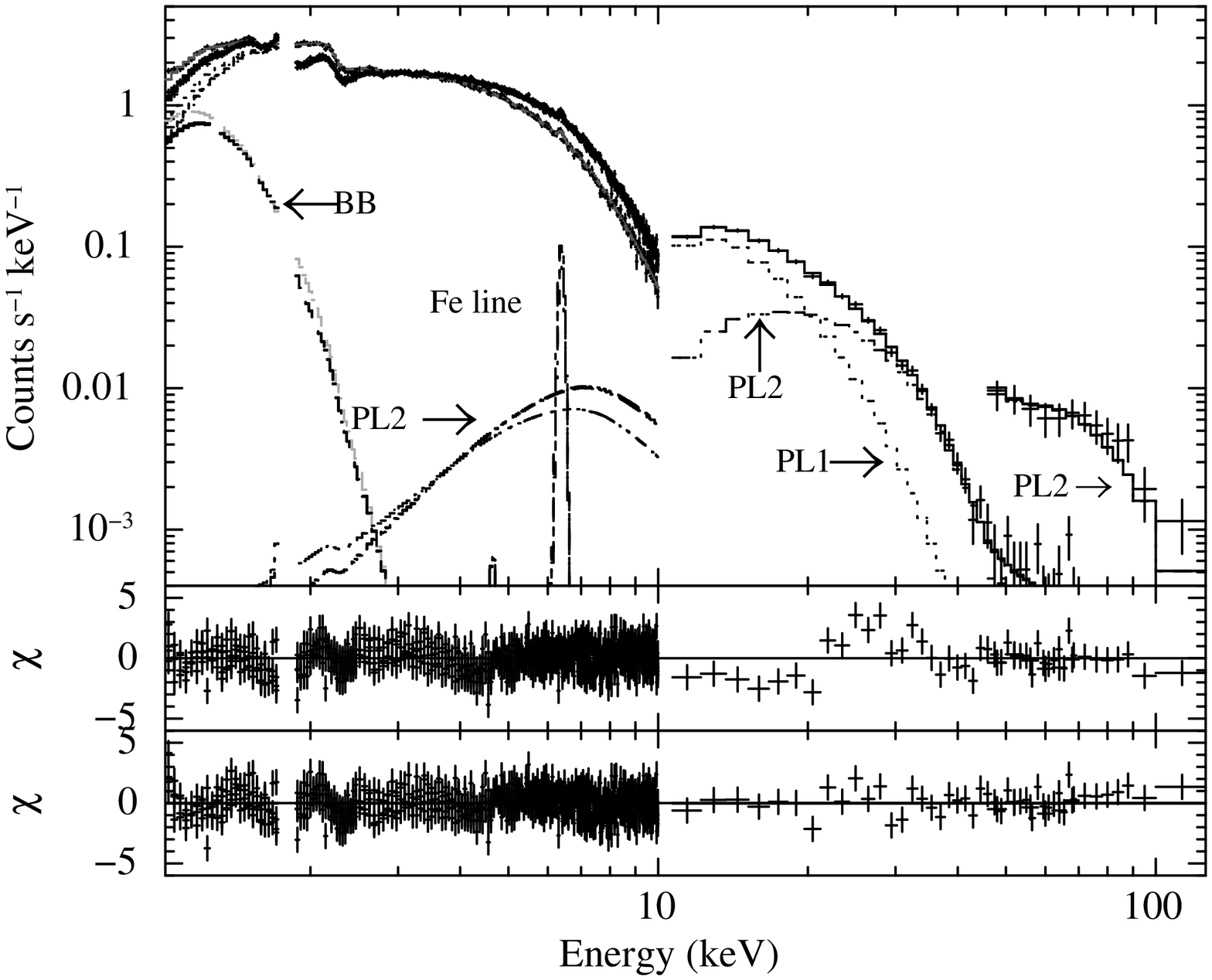}
\caption{Energy spectrum of A0535$+$262 obtained with the XIS, PIN, and GSO
detectors of the \textit{Suzaku} observation, along with the best-fit model
comprising a blackbody component (BB), an NPEX power-law continuum model, a
narrow iron line emission (Fe line), and a cyclotron resonance factor at $\sim
45$ keV. The negative and positive power laws are marked by PL1 and PL2,
respectively. The middle and bottom panels show the contributions of the
residuals to $\chi^{2}$ for each energy bin for the cutoff power-law continuum
model and the NPEX continuum model, respectively.}\label{a0535_spec}
\end{figure}

X-ray binary pulsars are known to have very strong surface magnetic fields in
the range of 10$^{12}$--10$^{13}$ Gauss. The surface magnetic field strengths
can be most accurately determined by measuring quantized electron cyclotron
resonance features corresponding to transition between adjacent Landau levels,
which are separated by $E_{a1} = 11.6 \times B_{12} (1+z)^{-1}$ (keV), where
$B_{12}$ is the magnetic field strength in the unit of 10$^{12}$ Gauss and $z$
is the gravitational redshift. In case of pulsars with magnetic field strength
$B_{12}$ in the range of 1 to 10, $E_{a1}$ falls in the hard X-ray energy
(10--100 keV) range. Therefore, detection of spectral absorption features at
this resonance, called Cyclotron Resonant Scattering Feature (CRSF) is a
powerful tool to accurately determine the pulsar magnetic fields. Using this
spectral feature, the surface magnetic fields of about 15 binary pulsars have
been accurately measured with several X-ray observatories such as
\textit{BeppoSAX}, \textit{RXTE}, \textit{Suzaku} and \textit{Integral}. 
(Coburn et al.\ 2002 and references therein). The energy spectrum of Be/X-ray
binary pulsar A0535$+$262 showing the presence of CRSF along with other model
components is shown in Fig.~\ref{a0535_spec}.

Most of these pulsars are in transient HMXB systems. Another very interesting
phenomenon in accreting X-ray pulsars is that several pulsars show luminosity
dependent changes in the cyclotron resonance energy. Some are 4U~0115$+$63,
A0535$+$262, X0331$+$53, Her~X-1  (Nakajima et al.\ 2006; Terada et al.\
2006; Nakajima, Mihara \& Makishima 2010 and references therein). In the case of 
transient
pulsars 4U0115$+$63 and X0331$+$53 (V0332$+$53), the cyclotron resonance energy
has been found to correlate negatively with the source luminosity. The anti
correlation between the X-ray luminosity and the cyclotron resonance energy is
understood as a result of a decrease in the accretion column height, in
response to a decrease in the mass accretion rate. Higher the luminosity, lower
in energy is the CRSF produced. This is because, higher luminosity implies
higher accretion rate and a larger height of the accretion column. The CRSF
feature is thus produced further away from the surface of the neutron star
(Nakajima et al. 2006) where the local dipole magnetic field is smaller.
However, the anti-correlation is not seen in all the sources and the reason for
this is not yet very clear.

\section{Time resolved spectroscopy and energy resolved timing}

Several of the X-ray observatories provide very good coverage in observation of
the transient sources during their outbursts, either due to very flexible
planning and maneuvering (\textit{RXTE}-PCA, \textit{Swift}-XRT) or due to the wide angle
coverage (\textit{Integral}-IBIS, \textit{Swift}-BAT). This has enabled detailed studies of the
timing and spectral properties over the long outburst periods of the HMXB
transients. The large photon collection area of the \textit{RXTE} and \textit{Suzaku} also allow
detailed pulse phase resolved spectroscopy or energy resolved pulse profile
studies and some important results obtained in the recent years are briefly
discussed here.

\subsection{Relation between spectral evolution and changes in the PDS}

In transient accreting X-ray binary pulsars which show large variation in the
rate of mass accretion during the X-ray outburst, it is found that the PDS
break frequency follows the variations of the X-ray flux. This represents the
changes of the magnetospheric radius with the mass accretion rate (Revnivstev
et al.\ 2009). The increase in the mass accretion rate results in decreasing
the size of the magnetosphere and hence the inner radius of the disk which
brings the system away from the corotation, so that the characteristic
frequency at the inner edge of the disk/flow increase. This has been verified
by examining the PDS of transient X-ray binary pulsars such as A0535$+$262,
4U~0115$+$63, V0332$+$53, KS~1947$+$300 (Revnistev et al.\ 2009; Reig 2008). It
was found in these systems that the break frequency in the PDS changes with the
X-ray luminosity as $f_{\rm b} \propto L_{\rm X}^{3/7}$ which agrees with the
standard theory of accretion onto magnetized compact objects.

Investigation of the timing and spectral variability of accreting Be/X-ray
binary pulsars during major outbursts inferred that the transient Be binaries
exhibit two branches in the colour-colour and colour-intensity diagrams. These
are the horizontal branch corresponding to a low-intensity state showing the
highest fractional rms, and the diagonal branch that corresponds to a
high-intensity state during which the source stays for about 75\% of the total
duration of the outburst. Though the power density spectra of Be/X-ray binary
pulsars are complex due to the peaks of the pulse period and its harmonics, the
aperiodic variability can generally be described with low number of Lorentzian
components (Reig 2008). Analyzing the colour-colour diagram and power density
spectra of 4U~0115$+$63, KS~1947$+$300, EXO~2030$+$375 and V0332$+$53, it was
found that the spectral-timing behaviour in Be/X-ray binaries shows the
existence of spectral branches in the colour-colour and colour-intensity
diagrams. At high and intermediate flux, the sources move along the diagonal
branch; at very low count rate the soft colour decreases, while the hard colour
remains fairly constant defining a horizontal branch. The low-intensity states
are found to be more variable in terms of fractional rms.

\subsection{Energy dependence of the pulse profiles}

\begin{figure}
\centering
\includegraphics[height=4.8in, width=3.0in, angle=-90]{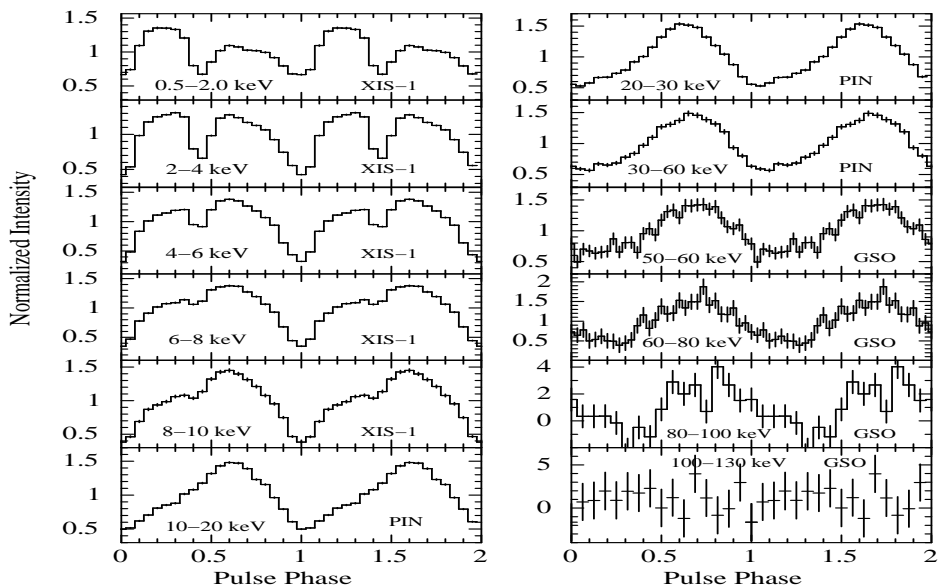}
\caption{Energy resolved pulse profiles of GRO~J1008-57. The presence/absence
of the dip-like structure in the 0.35-0.45 pulse-phase range can be seen.
The figure is taken from Naik et al.\ (2011a).}\label{J1008_pp}
\end{figure}

\begin{figure}
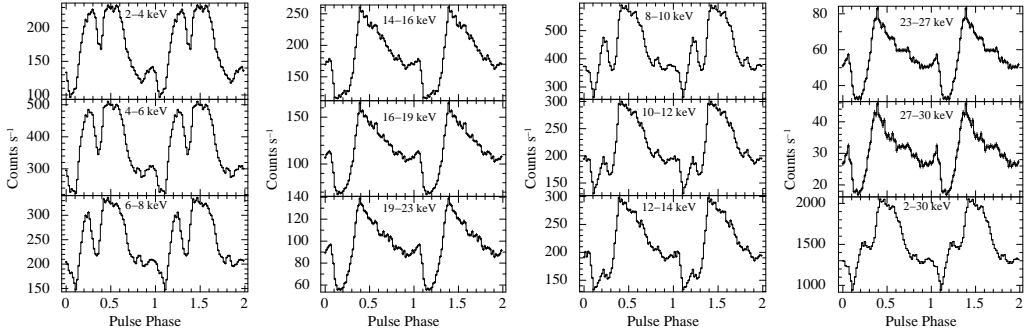

\centering
\includegraphics[width=4.3 cm, angle=-90]{Fg7a.ps}
\includegraphics[width=4.3 cm, angle=-90]{Fg7c.ps}
\includegraphics[width=4.3 cm, angle=-90]{Fg7b.ps}
\includegraphics[width=4.3 cm, angle=-90]{Fg7d.ps}
\caption{Energy dependent pulse profiles of GX 304-1 in different energy bands
observed with \textit{RXTE} during the peak phase of the outburst (2010 August
15$-$ MJD 55423). The figure is taken from Devasia et al.\
(2011b).}\label{gx304-1}
\end{figure}

Luminosity and energy dependence of pulse profiles are seen in several
transient HMXB pulsars. The pulse profiles of many transient sources show very
strong energy dependence, often a multiple-peaked profile at soft X-ray energy
bands ($\geq 8$ keV) and a single-peaked or double-peaked smooth profile at
hard X-rays (for example GRO~J1008$-$57; Fig.~\ref{J1008_pp}; Naik et al.\
2011a). Among some of the recent transients, the pulse profiles of GX~304$-$1
(Fig.~\ref{gx304-1}), 1A~1118$-$61, and A0535$+$262 were found to be strongly
energy dependent. Sometimes, the energy dependence also evolves during the
outbursts. An interesting feature that appeared from the energy resolved pulse
profiles of many pulsars is dip-like features in the pulse profiles. From pulse
phase resolved spectroscopy, the dips are found to be due to a partial
obscuration of the X-rays by matter that is phase locked to the neutron star.

\subsection{Pulse phase resolved spectroscopy, dips and absorption in accretion
stream}

\begin{figure}
\centering
\includegraphics[height=3.5in, width=3.0in, angle=-90]{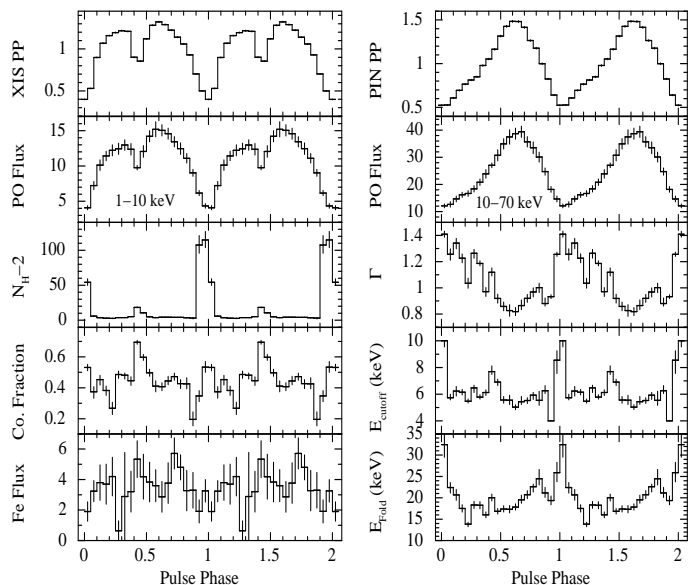}
\caption{Spectral parameters obtained from the pulse-phase-resolved
spectroscopy of \textit{Suzaku} observation of GRO~J1008$-$57. In the figure, the iron
line flux (Fe Flux), power-law flux (PO Flux) and NH2 are plotted in the units
of 10$^{-12}$, 10$^{-10}$ erg cm$^{-2}$ s$^{-1}$ and 10$^{22}$ atoms cm$^{-2}$,
respectively. The XIS and PIN pulse profiles are shown in the top left-hand and
top right-hand panels, respectively.}\label{gro_ph-rs}
\end{figure}

It has been found that the broad-band spectrum of transient HMXBs which show
dip-like features in the pulse profiles, is well described by partial covering
absorption model. There are cases where the phase averaged broad-band spectrum
of pulsars was well described by several continuum models. However, a partial
covering absorption model was found to be more suited over the power-law with
high energy cut-off or NPEX model to fit the pulse-phase resolved spectra. As
it is already described, in many cases, the power law with high energy cut-off,
NPEX and/or partial covering absorption models were all statistically fitting
well to the broad-band continuum spectra of transient pulsars such as
GRO~J1008$-$57 (Naik et al.\ 2011a), 1A~1118$-$61 (Devasia et al.\ 2011a;
Maitra et al. 2011), GX~304$-$1 (Devasia et al.\ 2011b). While investigating
the characteristics of the above pulsar at certain phases where dips or
dip-like features were present in the energy dependent pulse profiles, pulse
phase resolved spectroscopy showed that partial covering absorption model
described the pulsar spectrum well at almost entire pulse phase ranges.
Considering these results, the partial covering absorption model was preferred
while interpreting the results from pulse averaged spectroscopy. Pulse phase
resolved spectral analysis showed an increase in the value of absorption column
density of the partial covering component and variation in the covering
fraction during the dips or dip-like features in the pulse profile of the
pulsars. The changes in the values of column density ($N_{H2}$) and the
covering fraction during the dips in the pulse profiles naturally explain the
energy dependence of the pulse profiles in transient HMXB pulsars. The changes
in other spectral parameters such as iron emission line parameters, power law
photon index are also seen in dip pulse phases compared to other pulse
phases of the pulsars. A representative figure corresponding to the pulse phase
resolved spectroscopy of transient HMXB pulsar GRO~J1008$-$57 is shown in
Fig.~\ref{gro_ph-rs}.

\subsection{Pulse phase resolved measurements of the cyclotron lines}

Pulse phase resolved spectroscopy of the CRSFs in X-ray binary pulsars can
provide important information in understanding the combined effects of the
accretion flow geometry, the physics inside the accretion mound and radiation
transport in the highly magnetized plasma near the neutron star. To attempt
phase resolved spectroscopy of CRSFs in X-ray pulsars, sufficiently long
observations with detectors of excellent capability at hard X-rays are
required. Pulse phase dependence of the CRSF is known in a few sources (e.g. Vela
X-1: Kreykenbohm et al.\ 2002, 4U 1907$+$09: Rivers et al.\ 2010; GX 301$-$2:
Heindl et al.\ 2004). Recently, pulse phase resolved spectroscopy of the
CRSF in 1A~1118$-$61 was carried out using the \textit{Suzaku} observation of this
pulsar during its 2009 January outburst (Maitra et al.\ 2011). A remarkable
dependence of cyclotron energy and depth with pulse phase has been found in
1A~1118$-$61. In this source, the cyclotron energy and depth show a variation
of about 10 keV and by a factor 3, respectively, over the entire pulse phase. A
pulse phase dependence of the CRSF is expected because of the simple fact that
at different pulse phases the dipole geometry is being viewed at different
angles. This has importance in understanding the emission pattern, both
spectroscopic and geometric, from the accretion columns at the magnetic poles.
However, some deep and sharp pulse phase dependent features in the CRSF may
also imply a more complex underlying magnetic field structure.

\section{Super Fast X-ray Transient (SFXT) phenomena, link between supergiant
and Be XRBs?}

A new class of HMXBs has emerged in the last several years, which consists of a
compact star and a supergiant companion star. The remarkable feature of the
SFXTs is their fast transient nature and several models have been proposed to
explain this feature. Most of these binaries have wide orbits, very low X-ray
luminosity during the quiescent period, and X-ray pulsations have been detected
in few of these sources. One SFXT that stands out among all others is,
IGR~J16479$-$4514 (Jain, Paul \& Dutta 2009a) with the smallest known orbital period
of 3.32 d among the SFXTs and quiescent X-ray luminosity of $\sim 10^{34}$ erg
s$^{-1}$, about two orders of magnitude brighter than the other SFXTs in
quiescence. However, IGR J16479$-$4514 is much fainter than the persistent
HMXBs with similar orbital period and is thus thought to be a link between the
persistent HMXBs and the SFXTs while the SFXTs themselves are a link between
the systems with supergiant companions and the Be X-ray binaries. Regarding the
nature of the outbursts in the SFXTs, their duty cycle of the active period has
been measured in several systems using extensive monitoring programs (Sidoli et
al. 2008). There is evidence that the outbursts take place more frequently in
certain orbital phases compared to the rest of the binary periods (Jain, Paul \& Dutta
2009b) and this has implication for understanding of the SFXT mechanism.

\subsection{Optical/IR results during transient X-ray outbursts}

The optical companion of transient HMXB pulsars, specifically Be/X-ray pulsars
are B-emission (Be) spectral-type stars and are characterized by high
rotational velocities. These stars rotate at a speed close to critical limit so
that the surface gravity mostly balances with the centrifugal force around the
equator. The episodic equatorial mass loss in this type of stars forms a
circumstellar envelope, called a Be disk, that orbits around the star. The Be
stars show complicated line profiles containing an absorption component from
the photosphere and an emission component from the circumstellar disk. The
emission line profiles which reflect the nature of the circumstellar disk, show
many kinds of variabilities on the time scales from days to several years viz.
complete disappearance or truncation, evolution of the circumstellar disk,
changes in line profiles -- double peaked profiles etc. The change in line
profiles in Be/X-ray binaries was explained due to the change in electron
density in a region close to the Be star, triggered by uneven mass loss from
the optical companion.

During X-ray outbursts in transient Be/X-ray binary systems, there are several
occasions when extreme changes in line profiles of Be stars have been reported.
The disappearance of the H$\alpha$ emission line is generally used as a strong
indicator for disk loss in Be stars. Corbet, Smale \& Menzies (1986) found the H$\alpha$
profile to change from a shell profile to an absorption profile over a period
of four years in the Be/X-ray binary source 4U~1258-61 (GX~304-1). In case of
Be/X-ray binary V635~Cas, it was observed that the H$\alpha$ profile changed
from emission to absorption during 1997 February -- July (Negueruela et al.
2001). This was associated with a low photometric state in infrared magnitudes.
The circumstellar disk formed again after the disk loss episode within a period
of about 6 months. Disk loss in another transient HMXB pulsar A0535$+$262 was
inferred from a change in H$\alpha$ emission profile to absorption during 1997
October -- 1998 August (Haigh et al.\ 1999). It was also found that the
hydrogen lines like Br$\gamma$ in absorption on 1998 November, which is
associated with disk loss event. The H$\alpha$ line profile was found to show
remarkable variability during the giant X-ray outburst of A0535$+$262 in 2009
November -- December indicating the existence of a warped component (Moritani
et al.\ 2011). During 2011 February -- March X-ray outburst, a reduction in
photometric $JHK$ flux was also detected in A0535$+$262 binary system (Naik et
al.\ 2011c). The disappearance of the circumstellar material in Be/X-ray binary
X~Per was evident from the conversion of H$\alpha$ profile from emission to
absorption (Norton et al.\ 1991). They also found that the near-infrared flux
in $JHK$-bands was reduced, implying a common region of formation in
circumstellar plasma for infrared continuum and Balmer line emission. In
transient Be/X-ray binaries, the circumstellar disk loss phenomenon, related
changes in the emission line features in optical and near-infrared spectra and
the re-formation of the disk can occur over a period of a few days/weeks to
months/years. Therefore, frequent monitoring of the optical companion in the
HMXB system may detect several of such events during the transient X-ray
outbursts which can lead to better understanding of the cause of such events.

\subsection{Future with Astrosat}

The first Indian space astronomy mission Astrosat, due for launch in 2012 will
bring some qualitative changes in the study of some aspects of HMXB transients.
Similar to \textit{BeppoSAX} and \textit{Suzaku}, it will provide a wide energy coverage.
However, a much larger effective area of the LAXPC instrument (Paul 2009) in
the 20--80 keV band compared to the earlier and current missions will enable
very detailed studies of the cyclotron line features. In addition, Astrosat
will also provide unprecedented simultaneous multi-wavelength observations, in
three bands in the optical/UV region. This will enable some sensitive probe of
the X-ray reprocessing in the surrounding medium, especially the accretion
disk.

\section{Acknowledgments}

We would like to thank the editors, D.~J.\ Saikia and D.~A.\ Green, for
inviting us to write this article. We also thank Rajeev R. Jaiswal for
carefully reading of the manuscript. The research work at Physical Research
Laboratory is funded by the Department of Space, Government of India. This
research has made use of data obtained through the High Energy Astrophysics
Science Archive Research Center Online Service, provided by the NASA/Goddard
Space Flight Center.

\label{lastpage}
\end{document}